\documentclass{aa}
\usepackage{graphicx}
\usepackage{balance}
\usepackage{txfonts}
\usepackage{changes}
\usepackage{natbib}
\usepackage{hyperref}
\hypersetup{
  colorlinks   = true, 
  urlcolor     = blue, 
  linkcolor    = blue, 
  citecolor   = blue 
}
\usepackage{color}

\begin{document}

 \title{Photometry of the eclipsing cataclysmic variable \\ SDSS J152419.33+220920.0.\footnote{Photometry described in Table \ref{tab:photolog} is only available in electronic form at the CDS via anonymous ftp to cdsarc.u-strasbg.fr (130.79.128.5) or via \url{http://cdsweb.u-strasbg.fr/cgi-bin/qcat?J/A+A/}}}

   \author{R. Michel\inst{1}, J. Echevarr\'{\i}a\inst{2}and J.V. Hern\'andez Santisteban\inst{2}\inst{,3}}

   \institute{Observatorio Astron\'{o}mico Nacional, Instituto de Astronom\'{i}a, Universidad Nacional Aut\'onoma de M\'exico,
   Apartado Postal 877, 22830, Ensenada, B.C., Mexico\\
              \email{rmm@astrosen.unam.mx}
         \and
         Instituto de Astronom\'{i}a, Universidad Nacional Aut\'onoma de M\'exico, Apartado Postal 70-264, M\'exico, D.F., M\'exico  \\
             \email{jer@astro.unam.mx,jvhs1g12@soton.ac.uk}
          \and
         School of Physics \& Astronomy, University of Southampton, Highfield, Southampton SO17 1BJ, UK.
 }

   \date{Received --- ; accepted ---}

\authorrunning{Michel, Echevarr\'{\i}a \& Hern\'andez Santisteban}
\titlerunning{The eclipsing CV SDSS J1524}

    \abstract
  {}
   {We present new photometry of the faint (g $\sim19$ mag) and poorly studied cataclysmic variable SDSS J152419.33+220920.0,
    analyze its light curve and provide an accurate ephemeris for this system.}
   {Time-resolved CCD differential photometry was carried out using the 1.5m and 0.84m telescopes at the Observatorio Astron\'omico Nacional at San Pedro M\'artir.}
   {From time-resolved photometry of the system obtained during six nights (covering more than twenty primary eclipse cycles in more than three years),
   we show that this binary presents a strong primary and a weak secondary modulation. Our light curve analysis shows that only two fundamental frequencies are present, corresponding to the orbital period and a modulation with twice this frequency. We determine
   the accurate ephemeris of the system to be $HJD_{eclipse}~ =~2454967.6750(1)~+~0.06531866661(1)~E$. A double-hump orbital period modulation, a standing feature in several bounce-back systems at quiescence, is present at several epochs. However, we found no other evidence to support the hypothesis that this system belongs to the post-minimum orbital-period systems.}
   {}


   \maketitle
%

\section{Introduction}
\begin{figure}[h]
 \setlength{\unitlength}{1mm}
 \includegraphics[width=90mm,trim=2.6cm 0.1cm 1.9cm 1.1cm,clip=]{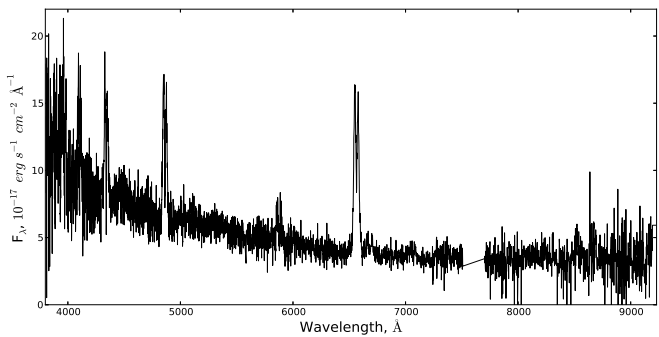}\\
 \caption{Spectrum of SDSS~1524. Data are taken from the SDSS database.}
 \label{fig:1}
\end{figure}

The discovery of the cataclysmic variable  SDSS J152419.33+220920.0 (hereafter SDSS~1524) was announced in the
seventh CV release of the Sloan Digital Sky Survey (SDSS) \citep{szk09}. The complete SDSS optical spectrum of this object is shown in Figure~\ref{fig:1}.
It shows a blue continuum with strong double-peaked Balmer emission lines, which is typical of high-inclination systems,
and, as Szkody and collaborators pointed out, a candidate for showing deep eclipses of the white dwarf by the secondary star.
No spectroscopic features from the donor star are detected in the red part of the spectrum.
The object, at quiescence, has a magnitude of g$\sim$19 mag; two outbursts have been detected by the
Catalina Real-Time Transient Survey (CRTS) \citep{dea09}. The first one\footnote{\url{http://nesssi.cacr.caltech.edu/catalina/20090329/903291210784144089p.html}} was reported first on March 30 2009 by the Variable Star Network \citep[VSNET, see][]{kat04}\footnote{VSNET-alert 11133}
and was subsequently confirmed by the Center for Backyard Astrophysics. The latter pointed out that the object was previously followed at quiescence for some weeks and
showed deep eclipses with a period of 0.065317 days\footnote{\url{http://cbastro.org/communications/news/messages/0635.html}}.
On March 31 2009 the AAVSO Cataclysmic Variable Network confirmed the presence of superhumps\footnote{cvnet-outburst/message/3029},
and of a 1.5~mag eclipse when the system was at magnitude 15. The second outburst on January 18 2012 was also detected by the CRTS and
by the VSNET\footnote{VSNET-alert 14121}. This outburst had a confirmed magnitude of $\sim$15.5 but no other observations have been reported.

An analysis of the 2009 outburst showed positive superhumps with a period of 0.06711 days \citep{kat09}. These authors also provided
ephemeris of the eclipses during outburst: Min(BJD)~=~2454921.5937(1) + 0.0653187(1) E.
\citet{gan09} included SDSS~1524 in a large study of faint newly discovered SDSS CVs, concluding mainly that the orbital
period distribution statistics have greatly improved thanks to the nearly 140 objects. This new
faint sample, not surprisingly, favours detection of short-orbital-period systems.
\citet{sou10}, based on two eclipses observed on 6 May 2009, using the WHT in service mode, derived the following ephemeris:
Min(HJD)~=~2454957.6499(1) + 0.06500(32)  E.
The light curve of their second eclipse shows evidence of an occultation of the hot spot and the white dwarf.

Since this object requires a comprehensive photometric analysis, we decided to observe it at different epochs and collect enough information to improve on its light curve, eclipses, and ephemeris. 
Thus, in this paper we present new time--resolved photometry of SDSS~1524 and the results of its light curve behaviour, as well as an improved ephemeris. 
We discuss the possible membership of this object to post--minimum-orbital-period systems.

\section{Observations}
\label{sec:observations}
\begin{figure}
\includegraphics[width=85mm,trim=-1.0cm 0.0cm 0.0cm 0.0cm,clip=]{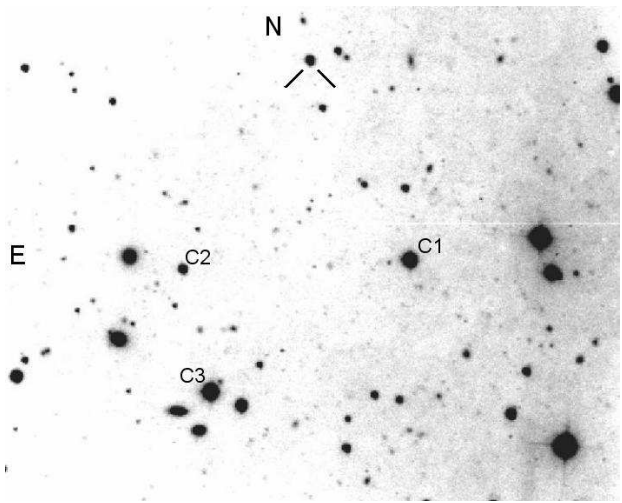}
 \label{fig:2}
 \caption{Finding chart for SDSS~1524. The field is $\sim4\times3.5$ arcmin.
The differential photometry was made with respect to star C1 (SDSS J152415.99+220804.0, g=17.77), while stars C2 (SDSS J152422.53+220753.5, g=19.23) and C3 (SDSS J152421.46+220705.7, g=16.97) 
were used as references to check that C1 was not a variable as well.}
 \end{figure}

We obtained white-light time--resolved photometry of SDSS~1524 during 16, 21 and 22 May 2009 
at the 1.5 m telescope of the Observatorio Astron\'omico Nacional at San Pedro M\'artir.
Additional observations were obtained with the 0.84m telescope on 20 May and 7 June 2011 and 8 August 2012.
In all runs we used the Blue-ESOPO CCD detector on a
$1024\times1024$ pixel configuration \citep{eea08}. The exposure times were between 100 and 120 sec.
In total, the object was monitored for $\sim 30$h and nearly 24 complete eclipses were measured.
Data reduction was performed with the IRAF\footnote {IRAF is distributed by the National Optical Observatories,
operated by the Association of Universities for Research in Astronomy, Inc., under
cooperative agreement with the National Science Foundation.} software.
The images were corrected for bias and were flat-fielded before aperture photometry was carried out.
An estimate of the uncertainty of the CCD photometry of SDSS~1524 is about 0.02 magnitude outside eclipse.
Differential photometry was obtained using star C1 as shown in Figure \ref{fig:2}.
The log of observations is presented in Table \ref{tab:photolog}.

\begin{table}
\caption{Log of photometric observations}
\label{tab:photolog}
\centering
\begin{tabular}{c c c c c}     
\hline\hline
Date  & HJD begin &HJD end    & No. of& Exposure \\
   ( UT )       &+2450000   &+2450000&   images &   Time  (sec)                      \\
    \hline
    May 16 2009& 4967.71141 & 4967.98379 & 181 & 100\\
    May 21 2009& 4972.91246 & 4973.00072 &  72 & 100 \\
    May 22 2009& 4973.64853 & 4973.99673 & 286 & 100\\
    May 20 2011& 5701.69035 & 5701.99199 & 187 & 120\\
    Jun 07 2011& 5719.66400 & 5719.92770 & 173 & 120\\
    Aug 08 2012& 6147.64344 & 6147.79328 & 113 & 100\\
   \hline\\
\end{tabular}
\end{table}

\section{Period analysis}
\label{sec:photoanalysis}

The photometric data of SDSS~1524 were analysed for periodicities using the discrete Fourier transform (DFT) code implemented in the \textit{Period04} \citep{len05} programme.
The power spectrum, shown in Figure \ref{fig:3}, shows the strongest peak at $f = 30.6191$  d$^{-1}$, because of a strong
primary eclipse and a shallow secondary eclipse. The primary eclipse alone gave a clear peak at $f = 15.3095593$  d$^{-1}$ corresponding to an orbital period of $P=94.06$ min., as shown in Figure \ref{fig:3}. The other peaks are harmonics {of this frequency. To fit the overall light curve, we used \textit{Period04} to calculate the phase and amplitude of its first 15 harmonics. The fit is shown as solid lines in Figures \ref{fig:4} and \ref{fig:5}.

\begin{figure}[h]
 \setlength{\unitlength}{1mm}
\includegraphics[width=90mm,clip=]{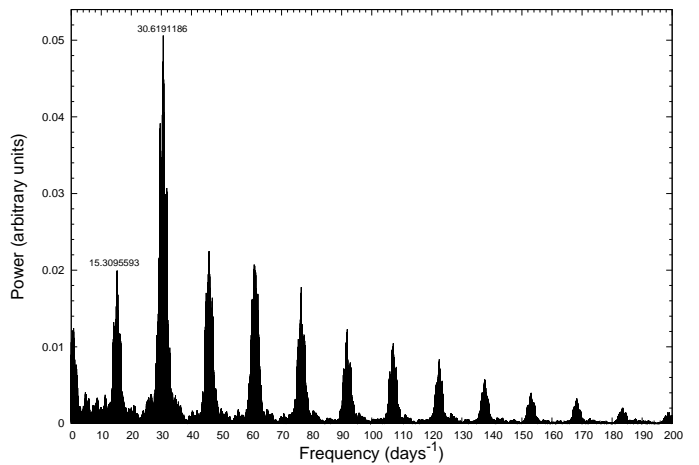}
 \caption{DFT analysis for SDSS~1524. The power spectrum shows the strongest peak at $f = 30.6191$ ~d$^{-1}$, while the orbital
 period correspond to $f = 15.3096$~d$^{-1}$. All other peaks are harmonics of these two fundamental frequencies.}
 \label{fig:3}
 \end{figure}

\section{Improved ephemeris}
\label{sec:ephemeris}

Shortly after the publication by \citet{szk09}, we started to observe SDSS~1524 and obtained a total of 24 orbital cycles
spanning more than three years. We have followed a consistent observational and reduction method throughout all epochs. Because of the symmetry of the eclipses, we fitted Gaussian profiles to determine the mid--eclipse timing.
This allowed us to calculate the improved ephemeris using only the data described in Section \ref{sec:observations},

\begin{equation}
HJD_{eclipse} = 2454967.6750(1) + 0.06531866661(1) E,
\end{equation}
where phase zero correspond to the mid-eclipse by the secondary star.

\section{Light curve behaviour}
\label{sec:light-curve-analysis}

\begin{figure}[h]
  \setlength{\unitlength}{1mm}
\includegraphics[height=100mm,width=90mm,trim=-0.5cm 0.0cm 0.0cm 0.0cm,clip=]{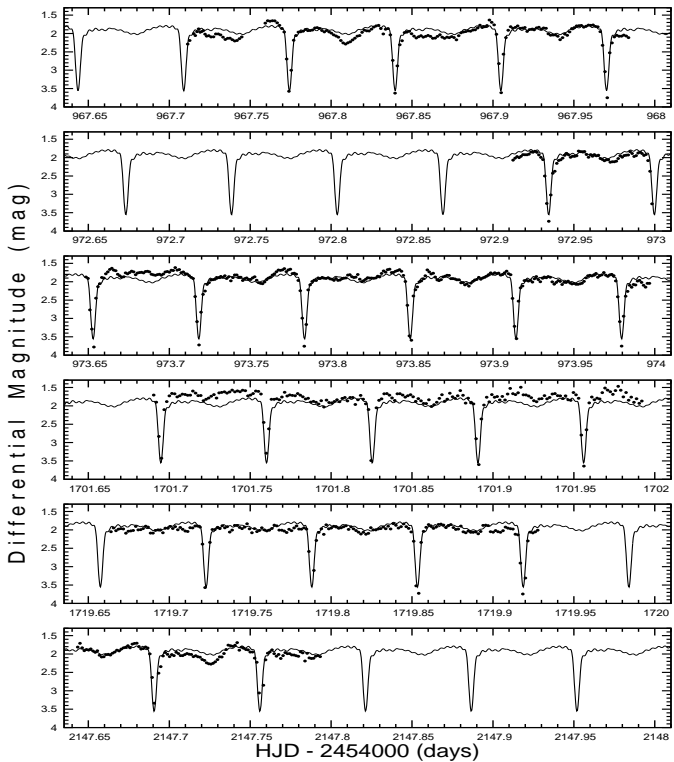}
 \caption{Differential photometry for SDSS~1524 with respect to star C1 ($circles$) and the DFT fit ($solid$ $line$) for all six runs (from top to bottom). See text for details of each observation.}
 \label{fig:4}
 \end{figure}

The photometric data are shown in Figure \ref{fig:4} together with the fit discussed in section \ref{sec:photoanalysis}.
The data are separated into the six observing runs for clarity, as presented in Table \ref{tab:photolog}. Although SDSS~1524 at first sight appears to have a regular photometric behaviour
with an almost constant difference of two magnitudes with respect to C1 and deep eclipses of more than 1.5 magnitudes, we point out that
the deep eclipses mask a variety of light curves, some of them similar to those observed in WZ~Sge \citep{pea98} (hereafter PEA98).
At some stages (first, third, and sixth run) the object shows a clear hump before primary eclipse in some cycles, with a decreased magnitude
at the egress, similar to that in PEA98 Figure 1, middle-panel and to the photometry reported in \citet{sou10}. This is typical of high-inclination 
systems with a bright spot, which are self-obscured by the accretion disc at this orbital phase \citep[e.g.][]{war95}.
At other stages (cycles in fourth and fifth runs), the system shows a double cycle variation with the orbital period, similar to those in
PEA98 (Figure~1 upper-panel), where the ingress and egress have comparable brightness. The difference in contrast between these two objects arises because 
the eclipses in WZ~Sge are not very deep; and is most certainly due to the difference in the inclination angle,
which in WZ~Sge is $\sim75^{\rm{o}}$ \citep{sma93}, while in SDSS~1524 it is $\sim83^{\rm{o}}$ \citep{sou10,sma10}.

The strength of the secondary minimum is variable from cycle to cycle, with a
highest depth of about 0.4-0.5 magnitudes (third cycle, first run) to almost an absence of it (e.g. fifth run). 
Its strength is not tied to the hump prior to eclipse, as show for example in the variety of the eclipses in the first run.
Its position in phase also appears to shift (almost up to $\phi= 0.6$). 
This is very clear in Figure \ref{fig:5}, where we have folded the data
with the orbital period. The small dots correspond to all observed points, while the open squares correspond to the data binned
in 0.02 phase intervals. The error bars represent the highest and lowest values at that interval. The solid line is the fit discussed in Section \ref{sec:photoanalysis}. While the binned data {\bf show} a minimum closer to phase 0.5, the fit favours a minimum closer to phase 0.6. This is probably due to the variation of brightness and location of a hot spot. The double cycle variation is more evident in Figure \ref{fig:5}. The secondary minimum is also present in the observations presented in \citet{sou10}.
The presence of the humps or the secondary minimum may be tied to the outbursts. This is evident, for example, in the first
2009 March outburst. Our observations in 2009 May show a strong activity in the light curves, while in our 2011 May run there seems to be a more quiet behaviour. Although we only have two nights around the 2012 outburst there is at least an indication of a lower activity prior to outburst than after.

\begin{figure}[h]
  \setlength{\unitlength}{1mm}
\includegraphics[height=60mm,width=90mm,trim=-0.5cm 0.0cm 0.0cm 0.0cm,clip=]{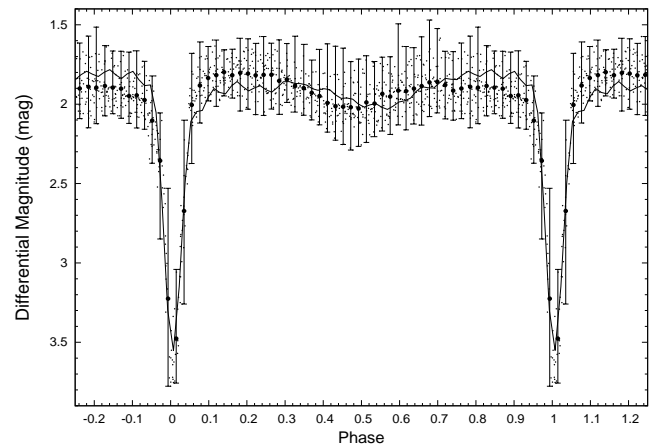}
 \caption{Folded photometry for SDSS~1524. The original data (dots) were binned in intervals of 0.02 in phase
 (open squares), the bars represent the extreme points in that phase interval. The solid line is the fit discussed in the text.}
 \label{fig:5}
 \end{figure}

\section{A bounce-back system?}

A large number of short-period cataclysmic variables have been found in
the Sloan Digital Sky Survey, confirming the prediction of an accumulation of systems close
to the orbital period minimum in the 80--90 minute range \citep{gan09}. Of these, a few
have been confirmed as moving away from this period minimum, the so-called bounce-back systems \citep{zea10,sav11}.
A great interest pertains to the observation of these binaries because they confirm the secular evolution of cataclysmic variables
at the short--end of the orbital period range (e.g. \citet{kab99} and references therein), and also because of
the physical study of brown dwarfs, here defined as stars that have lost mass up to the point where they can no longer
produce nuclear reactions, but are still semi--detached binaries and are losing enough mass to produce an accretion disc.

SDSS~1524 shows a double-hump orbital period modulation, a characteristic feature in several bounce--back systems at quiescence 
as described in \citet{zat09} and \citet{zea10}. These authors proposed that double-humps at quiescence might be an indication of a permanent 2:1 resonance 
in the accretion disc, which would require an extreme mass ratio ($q\leq0.1$). In SDSS~1524, this modulation has the strongest 
peak in our DFT analysis. Although the deep primary eclipses mask this modulation, it is clearly present in the binned data shown in 
Figure \ref{fig:5}. 
From their models of the light curve during primary eclipse, \citet{sou10} 
derived very similar results for SDSS~1524 and SDSS J115207.00+404947.8.

However, while SDSS J115207.00+404947.8 shows a shallow absorption in $H_{\beta}$ and other higher order 
Balmer lines, SDSS~ 1524 shows only a marginal absorption, and that only in $H_{\beta}$.  The presence of strong and broad 
Balmer lines in absorption (presumably coming from the white dwarf) are a strong feature in bounce-back systems. 
This is clearly the case in SDSS J103533.02+055158.3 \citep{sou06}. However, other candidates such as  
SDSS J123813.73-033933.0 \citep{zea06} and SDSS J080434.20+510349.2 \citep{szk06} show a less prominent white dwarf. 
In fact, SDSS J123813.73-033933.0 shows a variable contribution of the white dwarf at different epochs \citep{aea10}. 
It is possible that a similar behaviour might be found in SDSS~1524 and that the white dwarf will become visible. 
However, the high inclination of SDSS~1524 implies that we are viewing a more edge-on disc, which can hide 
the white dwarf contribution. More spectra are required and simultaneous photometry is strongly suggested to 
confirm if the brightness and appearance of the double-hump are correlated with spectral changes in this system. 
\citet{kat09} found the mass ratio from their superhump excess relationship to be $q=0.138$. Although their 
observations are scant and the actual light curve is not shown in their paper, their results and the fact that the 
secondary eclipse and hump is not always present argue strongly against a bounce-back object.

\begin{acknowledgements}

We would like to thank the anonymous referee for all his or her very useful corrections and comments, which greatly improved this paper. 
JE is indebted to DGAPA (Universidad Nacional Aut\'onoma de M\'exico) support, PAPIIT projects IN122409 and IN111713. JVHS was kindly 
supported by the MSc. Program at the IA-UNAM, CONACyT scholarship scheme and grant 129753.

\end{acknowledgements}
\bibliographystyle{aa}

\end{document}